%% file: main.tex
\documentclass[twocolumn]{aastex631}

\usepackage{verbatim}

\providecommand{\bjdtdb}{\ensuremath{\rm {BJD_{TDB}}}}
\providecommand{\feh}{\ensuremath{\left[{\rm Fe}/{\rm H}\right]}}
\providecommand{\teff}{\ensuremath{T_{\rm eff}}}

\providecommand{\So}{S$_{\oplus}$}

\providecommand{\cms}{cm\,s$^{-1}$}

\providecommand{\Kepler}{\textit{Kepler}}

\definecolor{pink}{rgb}{1,0.1,.6}

\shorttitle{Sorry they aren't in alphabetical order :(}
\shortauthors{Gilbert et al.}

\graphicspath{{./}{figures/}}

\begin{document}

\title{A Second Earth-Sized Planet in the Habitable Zone of the M Dwarf, TOI-700}

\correspondingauthor{Emily Gilbert}   
\email{emily.a.gilbert@jpl.nasa.gov}

\author[0000-0002-0388-8004]{Emily A. Gilbert}
\affiliation{Jet Propulsion Laboratory, California Institute of Technology, 4800 Oak Grove Drive, Pasadena, CA 91109, USA}

\author[0000-0001-7246-5438]{Andrew Vanderburg}
\affiliation{Department of Physics and Kavli Institute for Astrophysics and Space Research, Massachusetts Institute of Technology, Cambridge, MA 02139, USA}

\author[0000-0001-8812-0565]{Joseph E. Rodriguez}
\affiliation{Center for Data Intensive and Time Domain Astronomy, Department of Physics and Astronomy, Michigan State University, East Lansing, MI 48824, USA}

\author[0000-0001-5084-4269]{Benjamin J. Hord}
\affiliation{University of Maryland, College Park, MD 20742, USA}
\affiliation{NASA Goddard Space Flight Center, 8800 Greenbelt Road, Greenbelt, MD 20771, USA}

\author[0000-0001-8933-6878]{Matthew S. Clement}
\affiliation{Earth and Planets Laboratory, Carnegie Institution for Science, 5241 Broad Branch Road, NW, Washington, DC 20015, USA}
\affiliation{Johns Hopkins APL, 11100 Johns Hopkins Rd, Laurel, MD 20723, USA}

\author[0000-0001-7139-2724]{Thomas Barclay}
\affiliation{University of Maryland, Baltimore County, 1000 Hilltop Circle, Baltimore, MD 21250, USA}
\affiliation{NASA Goddard Space Flight Center, 8800 Greenbelt Road, Greenbelt, MD 20771, USA}

\author[0000-0003-1309-2904]{Elisa V. Quintana}
\affiliation{NASA Goddard Space Flight Center, 8800 Greenbelt Road, Greenbelt, MD 20771, USA}

\author[0000-0001-5347-7062]{Joshua E. Schlieder}
\affiliation{NASA Goddard Space Flight Center, 8800 Greenbelt Road, Greenbelt, MD 20771, USA}

\author[0000-0002-7084-0529]{Stephen~R.~Kane}
\affiliation{Department of Earth and Planetary Sciences, University of California, Riverside, CA 92521, USA}

\author[0000-0002-4715-9460]{Jon~M.~Jenkins}
\affiliation{NASA Ames Research Center, Moffett Field, CA 94035, USA}

\author[0000-0002-6778-7552]{Joseph~D.~Twicken}
\affiliation{NASA Ames Research Center, Moffett Field, CA 94035, USA}
\affiliation{SETI Institute, Mountain View, CA 94043, USA}

\author[0000-0001-9269-8060]{Michelle Kunimoto}
\affiliation{Department of Physics and Kavli Institute for Astrophysics and Space Research, Massachusetts Institute of Technology, Cambridge, MA 02139, USA}

\author[0000-0001-6763-6562]{Roland~Vanderspek}
\affiliation{Department of Physics and Kavli Institute for Astrophysics and Space Research, Massachusetts Institute of Technology, Cambridge, MA 02139, USA}

\author[0000-0001-6285-267X]{Giada N. Arney}
\affiliation{NASA Goddard Space Flight Center, 8800 Greenbelt Road, Greenbelt, MD 20771, USA}
\affiliation{GSFC Sellers Exoplanet Environments Collaboration, NASA Goddard Space Flight Center, Greenbelt, MD 20771}

\author[0000-0002-9003-484X]{David~Charbonneau}
\affiliation{Center for Astrophysics ${\rm \mid}$ Harvard {\rm \&} Smithsonian, 60 Garden Street, Cambridge, MA 02138, USA}

\author[0000-0002-3164-9086]{Maximilian~N.~G{\"u}nther} 
\affiliation{European Space Agency (ESA), European Space Research and Technology Centre (ESTEC), Keplerlaan 1, 2201 AZ Noordwijk, The Netherlands}
\affiliation{ESA Research Fellow}

\author[0000-0003-0918-7484]{Chelsea~X.~Huang}
\affiliation{University of Southern Queensland, West St, Darling Heights, Toowoomba, Queensland, 4350, Australia}

\author[0000-0002-8458-0588]{Giovanni Isopi}
\affiliation{Campo Catino Astronomical Observatory, Regione Lazio, Guarcino (FR), 03010 Italy}

\author[0000-0001-5347-7062]{Veselin B. Kostov}
\affiliation{NASA Goddard Space Flight Center, 8800 Greenbelt Road, Greenbelt, MD 20771, USA}
\affiliation{SETI Institute, Mountain View, CA 94043, USA}

\author[0000-0002-2607-138X]{Martti H. Kristiansen}
\affiliation{Brorfelde Observatory, Observator Gyldenkernes Vej 7, DK-4340 T\o{}ll\o{}se, Denmark} 

\author[0000-0001-9911-7388]{David~W.~Latham}
\affiliation{Center for Astrophysics $\mid$ Harvard \& Smithsonian, 60 Garden St, Cambridge, MA, 02138, USA}

\author{Franco Mallia}
\affiliation{Campo Catino Astronomical Observatory, Regione Lazio, Guarcino (FR), 03010 Italy}

\author[0000-0003-2008-1488]{Eric E. Mamajek}
\affiliation{Jet Propulsion Laboratory, California Institute of Technology, 4800 Oak Grove Drive, Pasadena, CA 91109, USA}

\author[0000-0002-4510-2268]{Ismael~Mireles}
\affiliation{Department of Physics and Astronomy, University of New Mexico, 210 Yale Blvd NE, Albuquerque, NM 87106, USA}

\author[0000-0002-8964-8377]{Samuel~N.~Quinn}
\affiliation{Center for Astrophysics $\mid$ Harvard \& Smithsonian, 60 Garden St, Cambridge, MA, 02138, USA}

\author[0000-0003-2058-6662]{George~R.~Ricker}
\affiliation{Department of Physics and Kavli Institute for Astrophysics and Space Research, Massachusetts Institute of Technology, Cambridge, MA 02139, USA}

\author[0000-0002-7382-0160]{Jack Schulte}
\affiliation{Center for Data Intensive and Time Domain Astronomy, Department of Physics and Astronomy, Michigan State University, East Lansing, MI 48824, USA}

\author[0000-0002-6892-6948]{S.~Seager}
\affiliation{Department of Physics and Kavli Institute for Astrophysics and Space Research, Massachusetts Institute of Technology, Cambridge, MA 02139, USA}
\affiliation{Department of Earth, Atmospheric and Planetary Sciences, Massachusetts Institute of Technology, Cambridge, MA 02139, USA}
\affiliation{Department of Aeronautics and Astronautics, MIT, 77 Massachusetts Avenue, Cambridge, MA 02139, USA}

\author[0000-0003-4471-1042]{Gabrielle Suissa}
\affiliation{Department of Astronomy and Astrobiology Program, University of Washington, Box 351580, Seattle, Washington 98195, USA}

\author[0000-0002-4265-047X]{Joshua~N.~Winn}
\affiliation{Department of Astrophysical Sciences, Princeton University, Princeton, NJ 08544, USA}

\author[0000-0002-1176-3391]{Allison Youngblood}
\affiliation{NASA Goddard Space Flight Center, 8800 Greenbelt Road, Greenbelt, MD 20771, USA}

\author[0000-0002-9428-1573]{Aldo Zapparata}
\affiliation{Campo Catino Astronomical Observatory, Regione Lazio, Guarcino (FR), 03010 Italy}

\begin{abstract}

We report the discovery of TOI-700 e, a 0.95 R$_\oplus$ planet residing in the Optimistic Habitable Zone (HZ) of its host star. This discovery was enabled by multiple years of monitoring from NASA's Transiting Exoplanet Survey Satellite (TESS) mission. The host star, TOI-700 (TIC 150428135), is a nearby (31.1\,pc), inactive, M2.5 dwarf ($V_{mag} = 13.15$). TOI-700 is already known to host three planets, including the small, HZ planet, TOI-700 d. The new planet has an orbital period of 27.8 days and, based on its radius (0.95 R$_\oplus$), it is likely rocky. TOI-700 was observed for 21 sectors over Years 1 and 3 of the TESS mission, including 10 sectors at 20-second cadence in Year 3. Using this full set of TESS data and additional follow-up observations, we identify, validate, and characterize  TOI-700 e. This discovery adds another world to the short list of small, HZ planets transiting nearby and bright host stars. Such systems, where the stars are bright enough that follow-up observations are possible to constrain planet masses and atmospheres using current and future facilities, are incredibly valuable. The presence of multiple small, HZ planets makes this system even more enticing for follow-up observations.

\end{abstract}


\section{Introduction} \label{sec:intro}

The field of exoplanets has come great lengths over the past several decades. In the early days of exoplanet science, the focus was largely on individual planet detections, showing that these worlds even exist and that astronomers are capable of detecting them. The first planets detected were truly alien: planets orbiting pulsars, blisteringly hot worlds orbiting their host stars in just a few days, or giant planets more massive than any seen in our Solar System \citep[e.g. ][]{Wolszczsan92, mayor95, bouchy05}. The nascent technology used in the early days of exoplanet discovery, however, prevented the detection of smaller planets, orbiting farther from their Main Sequence host stars. 

In the 2000s and 2010s, the introduction of new technology and facilities pushed the field closer to discovering Solar System analogs. The High Accuracy Radial-velocity Planet Searcher (HARPS, \citealt{HARPS}) used the Doppler method to reveal a huge population of sub-Neptune-mass exoplanets orbiting close to their stars \citep{mayor2011}. The launch of space missions like MOST \citep{most}, CoRoT \citep{corot}, and \Kepler\ \citep{borucki10} delivered the sensitivity to detect even smaller planets using the transit method.  The \Kepler\ mission in particular observed with high-enough sensitivity to routinely detect Earth-sized planets on temperate orbits around M dwarf stars, leading to the discovery of the first Earth-sized HZ planet \citep[Kepler 186-f, ][]{Quintana2014}. Analysis of the ensemble of \Kepler's planetary discoveries led to the realization that multi-planet systems are common, often have flat, coplanar architectures like that of our Solar System, and perhaps most profoundly, that HZ Earth-sized planets are common \citep{dressing13, petigura2013, dressing15, bryson2021}. Despite these breakthroughs, \Kepler\ targets were limited by the fact that most of \Kepler's planetary discoveries orbited stars too faint for detailed characterization using other telescopes on the ground or in space. Without knowledge of their bulk compositions and atmospheres, it was impossible to discern whether the habitable-zone Earth-sized planets \Kepler\ found in spades were truly Earth-like.

Fortunately, \Kepler's direct successor, the Transiting Exoplanet Survey Satellite (TESS) mission \citep{Ricker2015}, offers an opportunity to fill in the gaps in knowledge left by \Kepler. TESS operates using a different observing strategy from \textit{Kepler} - tiling the entire sky in 28-day long observations called sectors. TESS observes an area of sky that is 400 times greater than that surveyed in \Kepler's primary mission and is specifically targeting Earth's nearest stellar neighbors. One challenge resulting from TESS's observing strategy is that the shorter observation periods make it difficult to discover multi-planet systems with long-period planets. Fortunately, near the ecliptic poles, the TESS fields of view for each sector overlap, forming Continuous Viewing Zones (CVZs), enabling longer stares at targets. These longer stares allowed for the discovery and characterization of prominent TESS multi-planet systems such as L 98-59 \citep{kostov2019b} and TOI-270 \citep{vaneylen21}.

One notable discovery from TESS is the TOI-700 system of planets. TOI-700 is an M2.5 dwarf\footnote{TOI-700 has nearly identical colors and absolute magnitudes ($M_V$ = 10.64, $M_G$ = 9.60, $M_{Ks}$ = 6.17) as the M2.5V standard star \citep{Kirkpatrick1991} star GJ 226 ($M_V$ = 10.61, $M_G$ = 9.59, $M_{Ks}$ = 6.18).} that resides in TESS's Southern CVZ. In 2020, this system was shown to host 3 planets, including the first HZ, Earth-sized planet found using TESS data, TOI-700 d \citep{gilbert2020, Rodriguez2020}. Given the relative scarcity of small, temperate worlds currently known, TOI-700 d makes this system a particularly intriguing target of its own merit. Here, we introduce a fourth planet, TOI-700 e, orbiting between planets TOI-700 c and TOI-700 d, residing within the Optimistic HZ \citep{kopparapu2013}. There are only around a dozen small, HZ planets known to date, so this second HZ planet within a single system further emphasizes the importance of the TOI-700 system for future study.

In this paper, we first describe the observations taken by TESS in Years 1 and 3 of the mission, as well as a ground based observation of the new planet candidate (Section \ref{sec:observations}). Next, we vet the planet using difference imaging and statistically validate the planet using \texttt{vespa} in Section \ref{sec:vetting}. After showing that TOI-700 e is almost certainly a bona fide planet, we then fit all 4 planets in the system in Section \ref{sec:EFv2}, providing updated planet parameters from \citet{gilbert2020,Rodriguez2020} for planets b, c, and d. In Section \ref{sec:discussion}, we discuss the expected properties of planet e as well as prospects for follow up. We study the dynamical properties of the system and compare this system to other known multi-planet systems. Finally, we summarize the paper in Section \ref{sec:conclusion}.

\section{Observations} \label{sec:observations}

We obtained data from the ongoing TESS mission which observed TOI-700 in Years 1 and 3 of operation. These observations are described in detail in Section \ref{sec:tess-observations}. We also conducted ground-based observations to rule out astrophysical false positive scenarios for planet e with Campocatino Austral Observatory, which are described in Section \ref{sec:EB-check}. We also took advantage of the wealth of data collected by \citet{gilbert2020} and \citet{Rodriguez2020} which characterized the host star and validated planets b, c, and d.

\subsection{TESS Observations} \label{sec:tess-observations}
TOI-700 resides in the TESS Southern CVZ and was observed for extended stretches of TESS Years 1 and 3 of observations. During Year 1, TOI-700 was observed at 2-minute cadence during Sectors 1, 3, 4, 5, 6, 7, 8, 9, 10, 11, and 13 as a result of its inclusion in Guest Investigator Program Cycle 1 proposal G011180 (PI: C. Dressing).\footnote{Details of approved TESS Guest Investigator Programs may be found at \url{https://heasarc.gsfc.nasa.gov/docs/tess/approved-programs.html}.} TOI-700 was observed at short cadence in Year 3 of observations as a result of its inclusion in a number of Guest Investigator (GI) Programs: G03183 (PI: J. Rodriguez); G03264 (PI: J. Van Saders); G03278 (PI: A. Mayo); and G03068 (PI: D. Kipping). In particular, we present this work as a part of the approved GI program G03183 (PI: J. Rodriguez), which enabled TOI-700 to be observed at 20-second cadence in Year 3 of observations. In Year 3, TOI-700 was observed during Sectors 27, 28, 30, 31, and 33 - 38. Sectors missing in both years of TESS observations are a result of the target falling within the gaps of the detectors.

A search of the TESS Year 1 data revealed 3 transiting planet candidates. \citet{gilbert2020} used these data coupled with follow-up observations to validate the 3 candidates as bona-fide planets called TOI-700 b, c, and d. 
After the Year 3 data were collected, a search of the full dataset revealed an additional transiting planetary candidate orbiting interior to TOI-700 d, which we refer to in this paper as TOI-700 e. TOI-700 e transited 7 times during the first year of observations and another 7 times during the third year of observations (6 of which were uncontaminated by transits of other planets). The transits of TOI-700 e are somewhat shallower than those of TOI-700 d, which explains why it was not detected in the first year of observations alone, despite orbiting interior to the previously known TOI-700 d. 

The signal of TOI-700 e was first detected as a threshold crossing event (TCE) as a part of the TESS Science Processing Operations Center \citep[SPOC;][]{Jenkins2016} Sectors 1-39, multi-sector transit search conducted on 2021 July 31. The SPOC pipeline searches for planets using the Transiting Planet Search (TPS) software, which is an adaptive, noise-compensating matched filter \citep{Jenkins2002, Jenkins2010, Jenkins2020}. After detection by TPS, TOI-700 e passed all the diagnostic tests exercised and was reported in the Data Validation reports \citep{Twicken2018, Li2019}, including the difference image centroiding test, which located the source of this transit signature to within 6\farcs4 $\pm$ 3\farcs6 of TOI-700. After reviewing the DV reports and other diagnostics, the TESS Science Office issued an alert  upgrading this from a TCE to a TOI (TOI-700.04) on 2021 November 19 \citep{guerrero2021tess}.

In order to maximize the signal-to-noise ratio of the shallow transits of TOI-700 e, we generated custom light curves from the target pixel files (TPFs) for the target and used those light curves for our transit fits. We used 20-second cadence data when it was available (Year 3 data), which yielded higher precision than the 2-minute cadence data for the same time period. We reextracted the light curves from the TPFs using the systematics corrections described in \citet{Vanderburg2019}. In addition to decorrelating against the quaternion time series, we also decorrelated against the Co-trending Basis Vectors (CBVs) that are generated by the SPOC pipeline's Presearch Data Conditioning \citep[PDC, ][]{Stumpe2012, Stumpe2014,Smith2012} module. We elected not to decorrelate against the time series with the high-pass-filtered background levels as is sometimes done; including the background time series into our regression introduced artifacts into the final light curves. Finally, for each sector, we sampled 20 different apertures and chose the one which minimized photometric scatter to use for extracting the photometry. We use this light curve for our analysis going forward.

\subsection{Campocatino Austral Observatory}\label{sec:EB-check}

On 2021 December 27 UT, we observed an egress of the transit of TOI-700 e using the CAO (Campocatino Austral Observatory), situated at El Sauce Observatory in Chile and supported by the Obstech remote telescope hosting facility.\footnote{\url{https://obstech.cl/}} The CAO is a 61\,cm diameter, Planewave CDK24" telescope, and the observation was conducted by taking 360-second exposures using a clear filter. The goal was to identify or rule out the existence of any nearby eclipsing binary stars blended with TOI-700 in TESS's images that could have contributed the signal of TOI-700 e. Following the guidelines from the TESS Follow-up Observing Program Working Group (TFOP WG) Sub Group 1 (SG1) which conducts follow-up photometric observations of TESS planet candidates, we considered a target to be `cleared' (shown definitively not to be an eclipsing binary contaminant) if the NEBdepth (the depth of an eclipse needed to cause the transit signal) is greater than or equal to 5 times the root mean square scatter in the star's light curve (NEBdepth / RMS $>=$5). Likewise, if the NEBdepth is between 3 and 5 times the root mean square scatter, we consider the star ``likely cleared.'' With these data, we cleared or likely cleared 66 out of 86 stars within 2.5$^{\prime}$; no obvious nearby eclipsing binaries were detected.

\section{Vetting}\label{sec:vetting}

We have conducted several statistical analyses in order to ensure that the signals are real following the methods \citet{gilbert2020} used to validate TOI-700 b, c, and d. False positive planet candidates can come in several flavors including astrophysical signals (eclipsing binaries) hosted by nearby known blended stars, astrophysical signals from very close, and unknown blended stars, and instrumental artifacts. Validating planets is defined as retiring almost all of these potential false positive scenarios so that the probability that the transit signal is a false positive is very low (typically below $10^{-2}$). In this section, we perform tests to assess the likelihood that the signal from the new candidate is due to any of these sources. In particular, we performed extensive consistency tests to determine that the signal is not driven by a handful of ill-timed instrumental artifacts (Section \ref{sec:avmodshift}). We also used difference imaging (Section \ref{sec:dave}) to study the signal from the star at the pixel level in order to rule out eclipsing binaries from stars more than a few arcseconds away and discern whether or not the transits occur on target. Finally, we employed \texttt{vespa} (Section \ref{sec:vespa}) to determine the astrophysical false-positive probability for the planet and assess the likelihood that the signal is caused by very closely blended companions.

\subsection{Vetting for Instrumental Artifacts}\label{sec:avmodshift}

Small planets are notoriously difficult to detect due to their shallow transits and correspondingly low SNR. This was especially true during the \textit{Kepler} mission, but is also a known problem for the TESS mission where numerous low-Multiple Event Statistic (MES, a proxy for SNR reported by TPS) TOIs have later been found to be false alarms due to instrumental systematics. In such instances, it is generally best practice to confirm planet candidates using an independent facility as was done for TOI-700 d with \textit{Spitzer} \citep{Rodriguez2020}. Unfortunately, \textit{Spitzer} was decommissioned in early 2020, and is no longer available for planet follow up. Moreover, the CHEOPS \citep{benz2021} spacecraft, which has shouldered some of the observational burden left by \textit{Spitzer}'s decommissioning, cannot observe TOI-700 because the star is too far south from the ecliptic plane.

In the absence of available spacecraft for follow up, we are forced to instead rely solely on the TESS data in order to validate TOI-700 e. The MES of this detection is 7.2, which is barely above the customary detection threshold of 7.1 \citep{guerrero2021tess}. We therefore performed an analysis to both measure a more accurate signal-to-noise ratio than the estimate from the MES using an optimized transit shape and every transit for the signal. Our SNR calculation also incorporated red noise in order to assess how vulnerable we are to instrumental effects. We did this using a matched filter approach; we took the best-fit transit model from our \texttt{exofast} global analysis described in Section \ref{sec:EFv2} and convolved it with the light curve of TOI-700. We  subtracted the best fit models for planets b, c, and d prior to performing the convolution and were able to recover 2 additional transits that previously overlapped with transits of other planets in the system. This analysis resulted in a signal from TOI-700 e which produced a strong peak in the matched filter response when folded on the expected period and transit epoch of the signal. Compared to the scatter of the matched filter response out-of-transits, we find that TOI-700 e's signal has a significance of $10.05\sigma$. While signals with significance greater than $10\sigma$ are generally unlikely to be instrumental artifacts, TOI-700 e just barely surpasses that threshold. N.b., this analysis is  equivalent to the "mod-shift" test used by the \textit{Kepler} Robovetter software and in the \texttt{DAVE} package. 

\begin{figure*}
    \centering
    \includegraphics[width=\textwidth]{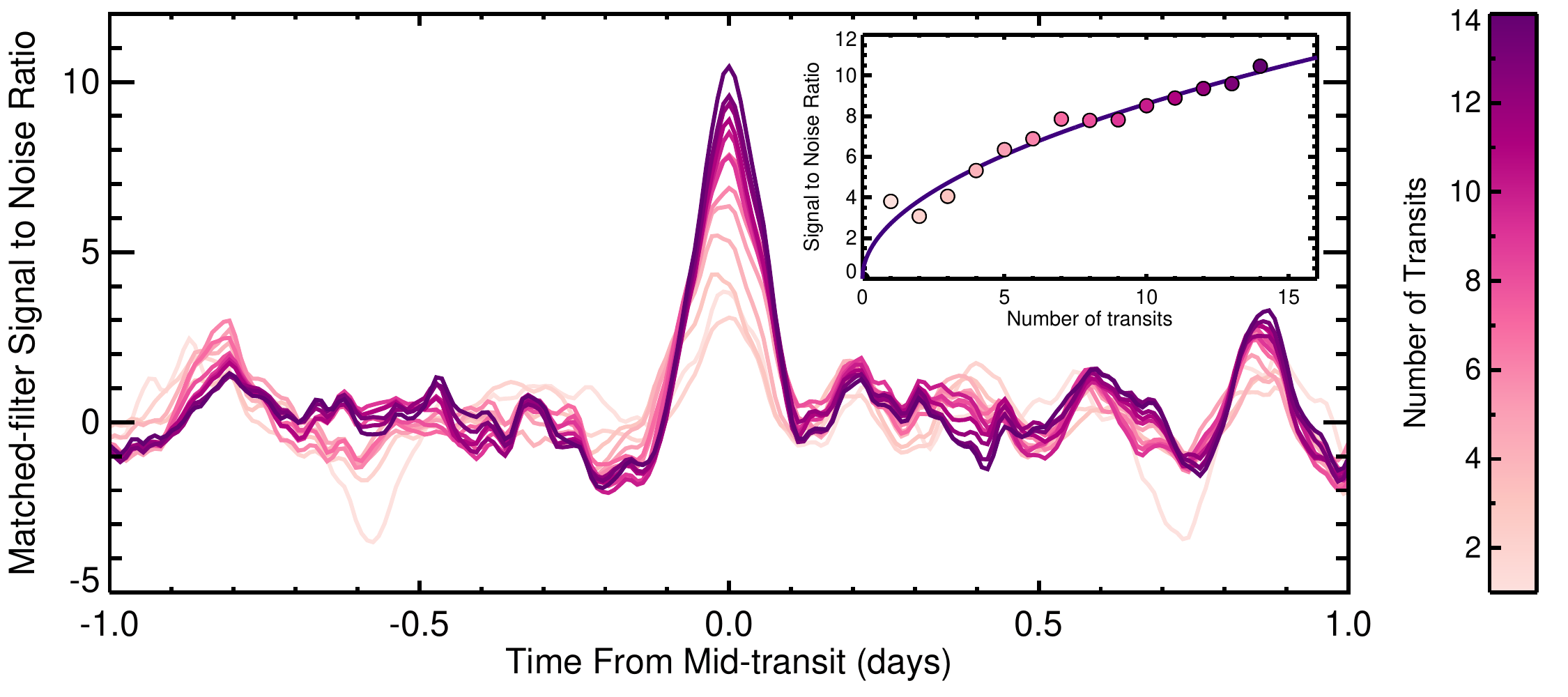}
    \caption{Signal-to-noise of TOI-700 e's transit detection as a function of number of observed transits. The main panel shows matched-filter response functions of a filter shaped like TOI-700 e's transit convolved with the TOI-700 light curve. The individual curves show the response of the light curve as additional data was collected. Curves with only a small number of transits (light pink) show a relatively weak detection of the planet's signal, while curves with many observed transits show a convincing detection, with greater than $10\sigma$ significance. The inset shows the peak signal-to-noise ratio of the planet's signal as a function of the number of observed transits. The purple curve is a best-fit curve with a square root power law, which describes the growth of the signal strength well, as expected for an astrophysical transit signal.  The consistent growth of the signal-to-noise as additional data was collected is evidence that the signal of TOI-700 e is not dominated by occasional instrumental artifacts, and instead is caused by a periodic signal with a period of 27.8 days. We note that in the parlance of \Kepler\ exoplanet detection literature, this type of metric is known as a ``mod-shift'' test.}
    \label{fig:e-SNR}
\end{figure*}

\begin{figure*}
    \centering
    \includegraphics[width=\textwidth]{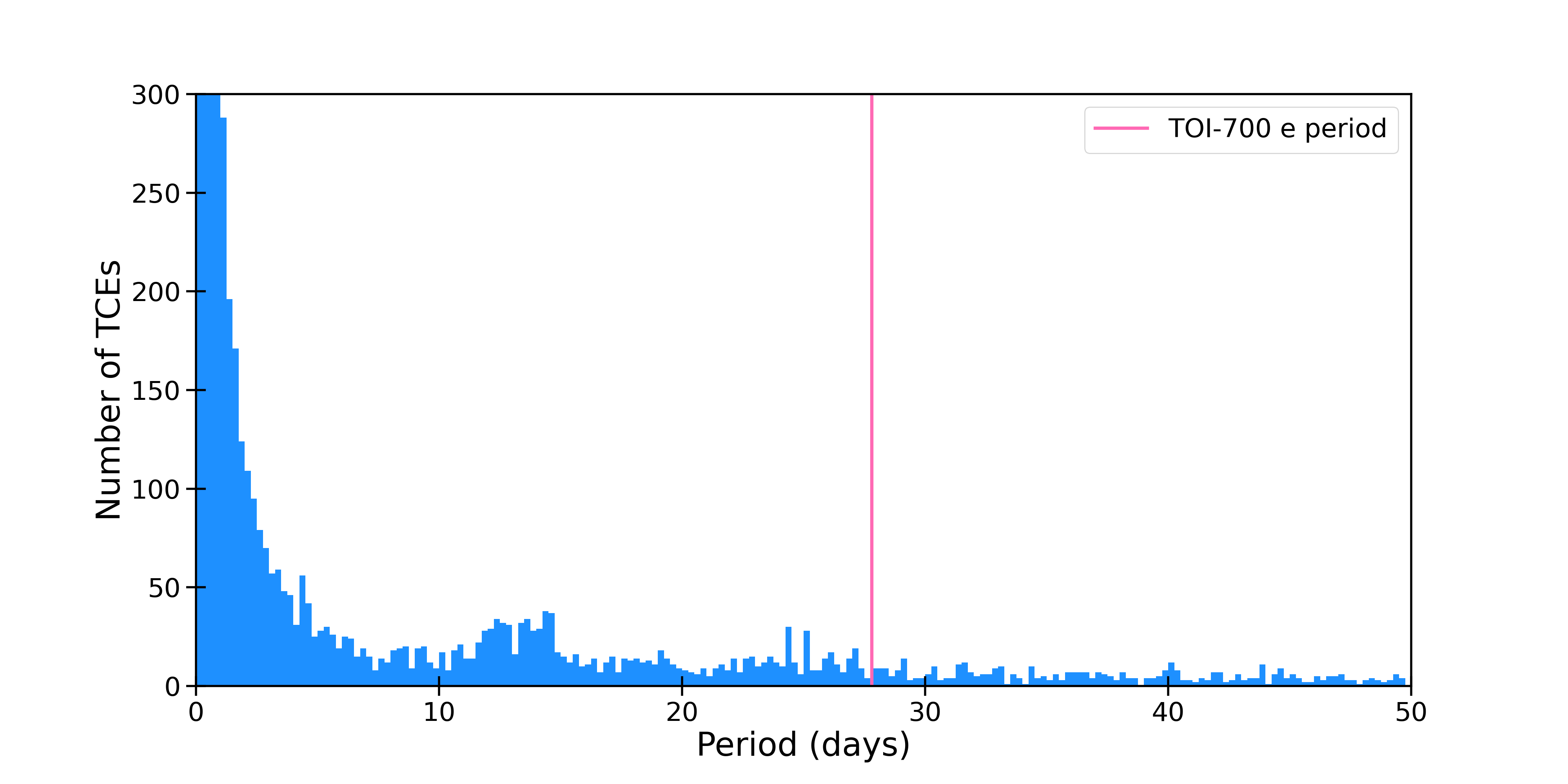}
    \caption{The distribution of periods of the TCEs shows that there is no significant pile up at the period of TOI-700 e (27.8 days) in this cutaway figure. The lack of overdensity at this period implies that there is no instrumental effect inducing planet signals at the period of TOI-700 e.}
    \label{fig:hist}
\end{figure*}

We also performed this same analysis on subsets of the data to test the growth of the signal as new transits were observed by TESS. We started by only analyzing data up to the first transit, then iteratively added each subsequent transit of planet e. We then plotted the significance of the peak in the matched filter response as a function of observed transit number. We found that the significance grows over time, and increases with each additional observed transit (see Figure \ref{fig:e-SNR}). This growth is consistent with a square-root curve, as would be theoretically expected for a real, periodic signal such as a planet transit. If the signal were caused by instrumental artifacts, they must be periodic with a 27.8 day period and each event must be relatively consistent in amplitude; if the signal were not repeating periodically with consistent amplitude, we would not see the growth in significance in this manner. We also checked the distribution of all TCEs detected by TESS to see if similar signals are present in other stars, which would imply an instrumental origin for both those signals and TOI-700 e. There is no significant overdensity of detected TOI or TCE signals with a period identical to 27.8 days, see Figure \ref{fig:hist}. These results, coupled the 14 transits of TOI-700 e observed by TESS spread over three years, strongly disfavor a signal of instrumental origin, thus implying that TOI-700 e's signal is indeed astrophysical.

\subsection{Difference Imaging}\label{sec:dave}

A common type of false positive is a nearby eclipsing binary (NEB) that contaminates the aperture pixels, causing the target light curve to show a transit-like signal. We used difference imaging in order to assess the evidence for this sort of contamination in the TESS pixel-level data.

We first employed \texttt{DAVE} (Discovery and Vetting of Exoplanets) \citep{Kostov2019} to study the potential sources of contamination. However, the combination of low SNR and long period of of the planet makes robust vetting with the \texttt{DAVE} pipeline unreliable for this target. Given there is at most one transit per sector, both the photocenter and the light curve modules of DAVE are vulnerable to cross-sector systematic effects. For example, non-astrophysical depth differences between different sectors would mislead the modshift module in interpreting two consecutive transits as exhibiting odd-even differences and flagging the planet candidate as a false positive. Without a definitive assessment from \texttt{DAVE}, we went on to use additional difference imaging to further vet this new planet candidate.

We then used the SPOC difference image centroid offsets \citep{Twicken2018}, which are averaged over all sectors in the Sectors 1-39 transit search in order to confirm that the transit occurred on target (within 6\farcs4 $\pm$ 3\farcs6 of TOI-700). No other stars bright enough to cause the transit signal of TOI-700 e lie within this region of sky, so this result implies that TOI-700 e's signal must be coming from either TOI-700 itself, or an unknown star blended with TOI-700. We verified this result using the difference imaging technique from \citet{Bryson2013} as implemented in the Python package \texttt{tess-plots}.\footnote{https://github.com/mkunimoto/TESS-plots} This method also confirmed that the transit occurred on target. The TESS Data Validation Reports also provide odd/even transits, and we inspected these and confirm that the depths match (consistent at the 0.83 sigma level), indicating that we are not seeing secondary eclipses of the planet candidate.

\subsection{Statistical vetting with \texttt{vespa}}\label{sec:vespa}
Although follow-up observations can rule out portions of parameter space where astrophysical false positives may exist, these observations are incomplete and do not yet definitively rule out the possibility that TOI-700 e's signal is an astrophysical false positive. Therefore, we supplemented our follow-up observations with a statistical analysis of the remaining likelihood of astrophysical false positive scenarios using the publicly-available software package, \texttt{vespa} \citep{2012ApJ...761....6M, 2015ascl.soft03011M} in order to establish the planetary nature of this signal.

\texttt{vespa} compares the transit signal to a number of astrophysical false-positives including an unblended eclipsing binary (EB), a blended background EB (BEB), a hierarchical companion EB, and EB scenarios with a double-period. The software assumes priors on the binary star occurrence rate from \cite{raghavan2010survey}, direction-specific star counts from \cite{girardi2005star}, and planet occurrence rates from \cite{2012ApJ...761....6M} before forward-modeling simulated populations for each category of false positive scenario. The known eclipse depth, duration, and ingress duration for these simulated populations are then compared against photometric data and planetary and orbital parameters provided by the user to calculate the likelihood that the target signal is well-described by one of these false positive populations. 

We ran \texttt{vespa} using the SPOC-produced light curves, which we phase-folded using the median values of the posteriors for the period and mid-transit time. The transits of the other planets in the TOI-700 system were subtracted from the light curve for this analysis. We included the maximum possible secondary depth of phase-folded features calculated by \texttt{DAVE} (Section \ref{sec:dave}) and the SOAR I-band contrast curve \citep[][]{gilbert2020} as observational constraints when calculating the false positive probability of the TOI-700 e signal. By default, \texttt{vespa} simulates the background starfield within 1 square degree of the target, but we dictated that the maximum aperture radius interior to which the signal must be produced be $42^{\prime\prime}$, which is the radius of 2 TESS pixels, and the maximum size of the aperture that the SPOC pipeline used to extract the light curves. \texttt{vespa} assumes that the transit-like signal is on target, which has been confirmed in this case by follow-up observations outlined in Section \ref{sec:observations} and the centroid analysis performed in Section \ref{sec:dave}.

Using these inputs, we calculate the false positive probability (FPP) of the TOI-700 e signal to be 2.74$\times$10$^{-4}$. All false positive scenarios have probabilities $\lesssim$10$^{-4}$, strongly disfavoring them over the planet scenario. Since the overall FPP $\ll$0.01, this signal can be considered to be statistically validated by \texttt{vespa}. Furthermore, this FPP does not account for the fact that the TOI-700 e signal is a part of a multiplanet system. \cite{Lissauer2012, Lissauer2014} demonstrated that false positives are less likely in multiplanet systems and that the FPPs calculated without accounting for this fact should be treated as upper limits. An analysis of TESS multiplanet systems indicates that this ``multiplicity boost" may reduce these FPPs by a factor of $\sim$50$\times$ \citep{guerrero2021tess}, further reducing the FPP of this planet. Given these results showing that astrophysical false positives are highly unlikely, and our analysis showing that the signal is not consistent with having an instrumental origin, we therefore state that TOI-700 e is almost certainly a bona fide planet.

\section{Transit Fits with EXOFASTv2 }\label{sec:EFv2}

\begin{figure*}[ht] 
   \centering
   \includegraphics[width=0.5\linewidth, angle =0, trim = 0 0 0 0 ]{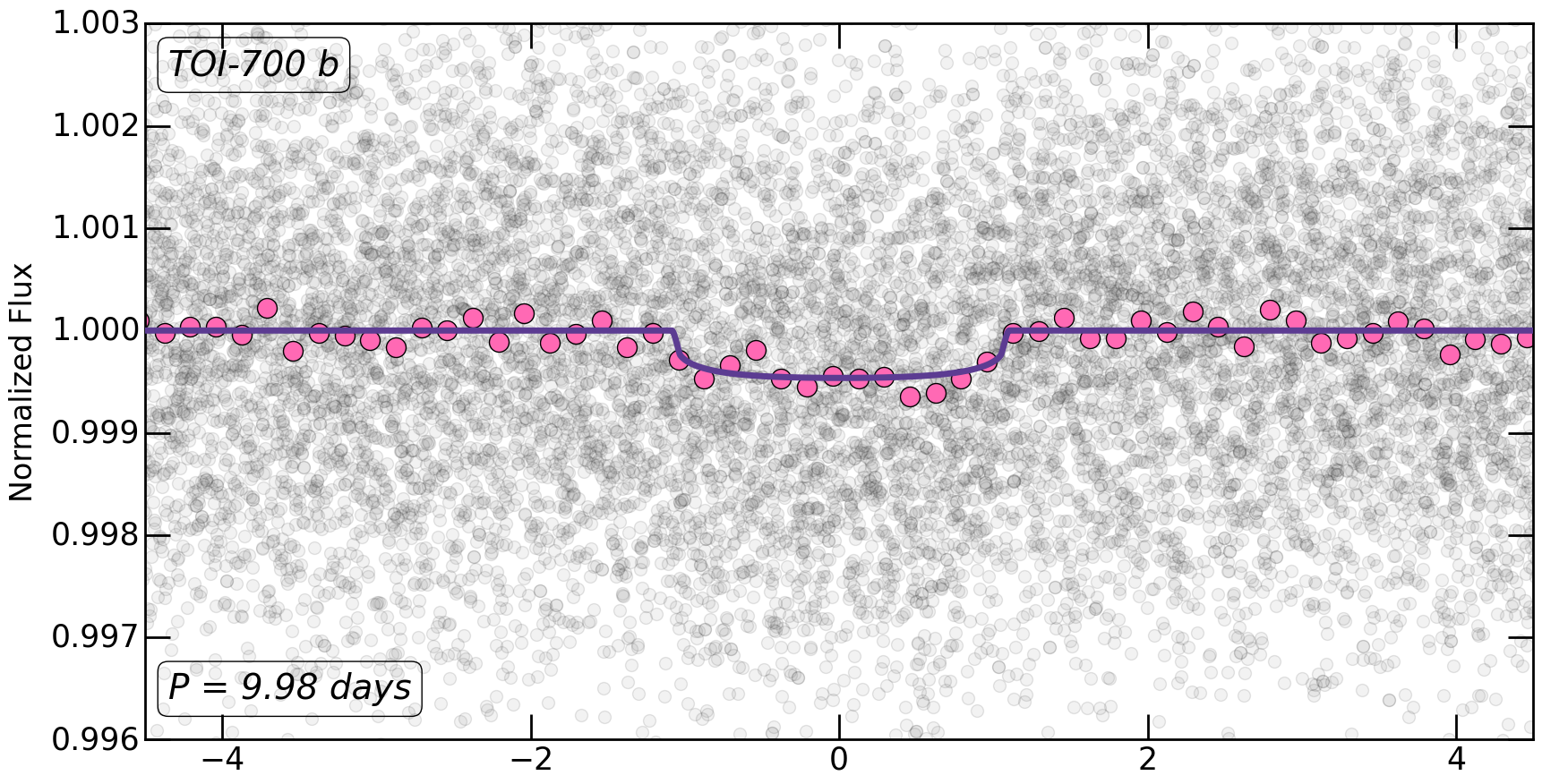}\includegraphics[width=0.5\linewidth, angle =0, trim = 0 0 0 0 ]{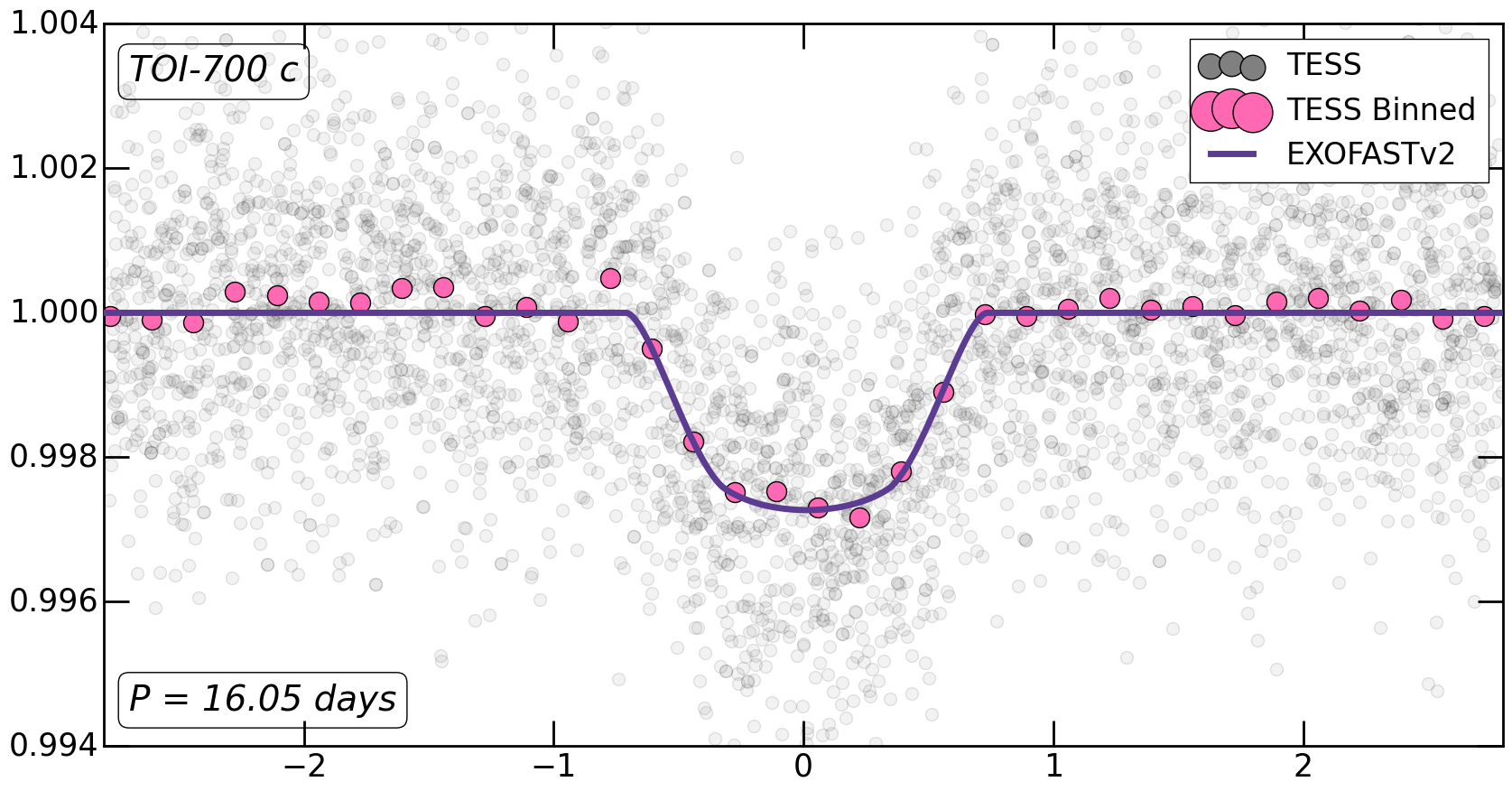}\\ 
   \includegraphics[width=0.5\linewidth, angle =0, trim = 0 0 0 0 ]{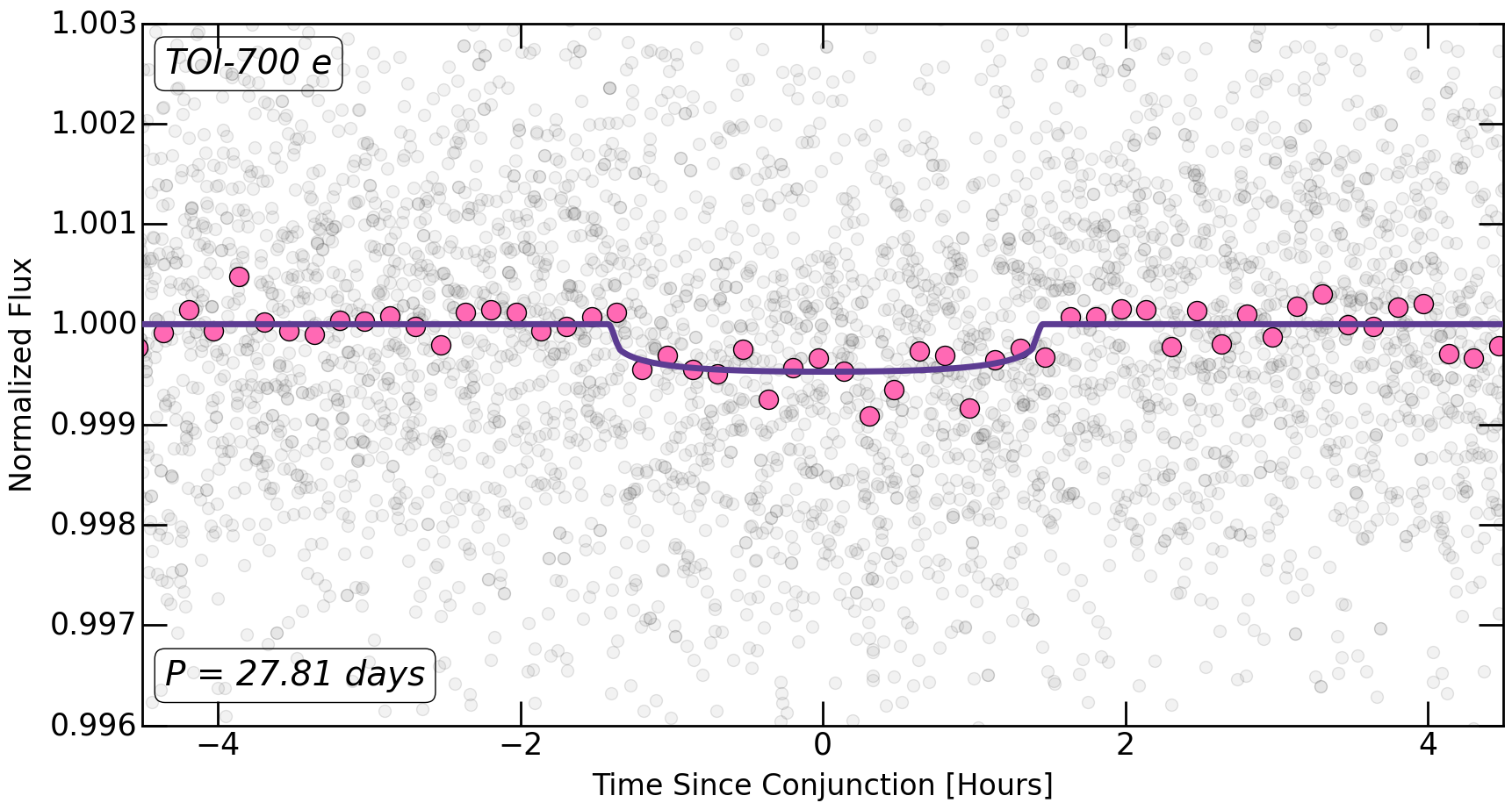}\includegraphics[width=0.5\linewidth, angle =0, trim = 0 0 0 0 ]{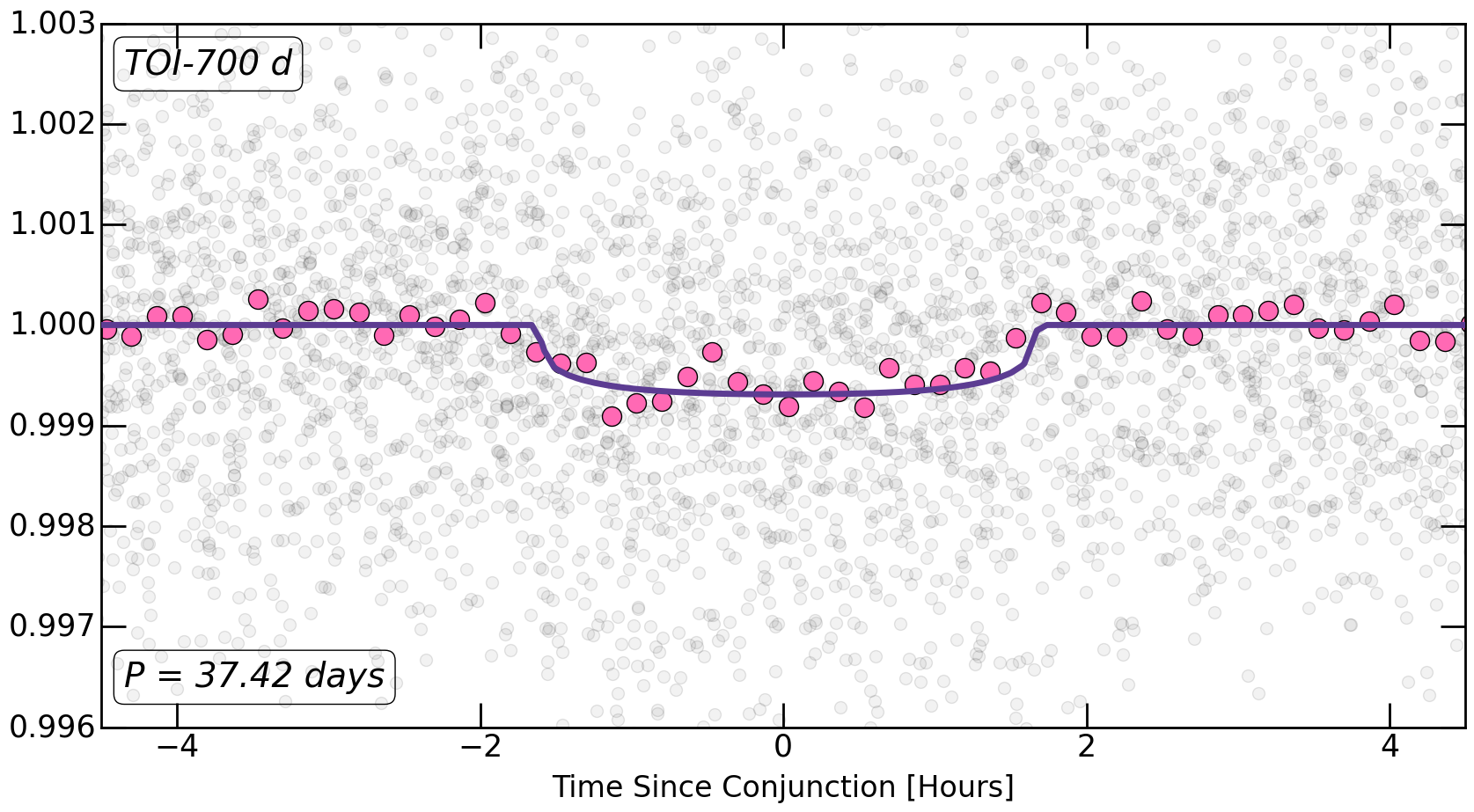}\\ 
\caption{Here we present the light curves for the TOI-700 planets phase folded to each of their orbital periods. The gray dots are the TESS 2-minute cadence data, the pink dots are the data binned to 10-minute cadence, and the purple lines show our transit fit generated using EXOFASTv2.} 
   \label{fig:LC}
\end{figure*}

To refine our understanding of the 3 known planets in the TOI-700 system and characterize the newly validated TOI-700 e, we globally fit the available photometric observations using the ExofastV2 software package \citep{Eastman:2019}. Our strategy generally followed that presented in Section 3 of \citet{Rodriguez2020}. We adopted the same priors on M$_{star}$ and R$_{star}$, $0.419\pm0.021$ R$_\odot$ and $0.417\pm0.021$ M$_\odot$, that were determined from the absolute $K_{s}$-mag relations from \citet{Mann2015} since stellar isochrones are unreliable for low-mass host stars. Priors on the \teff\ and \feh\ (3460 $\pm$ 65 K \& -0.07 $\pm$ 0.11 dex) were also included from spectroscopic analysis from the Southern Astrophysical Research (SOAR) telescope (see Section 3.1 of \citealp{gilbert2020}). We included the TOI-700 c follow-up transit from the Las Cumbres Observatory and the 2 \textit{Spitzer} follow-up transits of TOI-700 d in Channel 2 on the \textit{Spitzer} InfraRed Array Camera (IRAC, \citealp{Fazio:2004}) observed on 2019 October 22 UT and 2020 January 05 UT (Program 14314, PI: Vanderburg). See Section 2.2 and 2.3 from \citet{Rodriguez2020} for the data reduction and analysis of the LCO and \textit{Spitzer} observations. We included both the primary mission (Year 1) and extended mission (Year 3) observations of TOI-700 from TESS. During the extended mission of TESS, TOI-700 was placed on 20-second cadence target list; these observations yielded higher photometric precision than the 2-minute cadence data and allowed for a more detailed study of the system. For our global analysis, we binned the 20-second cadence data to 2 minutes and included the primary mission 2-minute cadence light curve. The results of these fits can be seen in Figure \ref{fig:LC}, Table \ref{tab:stellar-params} for the stellar parameters, and Table \ref{tab:planet-params} for the planetary parameters. Table \ref{tab:planet-params} contains both updated parameters for planets b, c, and d, as well as new parameters for planet e. As a result of updated data processing and background subtraction, planets b, c, and d have slightly smaller radii than have been previously reported. We find planet e to be a 0.95 R$_\oplus$ planet on a 27.8 day period orbit. This puts the planet within TOI-700's Optimistic HZ \citep{kopparapu2013}, receiving 1.27$\times$ Earth's insolation flux (\So).

\input{updated-exofast-table}

\input{planet-params}

\section{Discussion} \label{sec:discussion}

Here we describe the potential properties of TOI-700 e and evaluate the system as a whole. We discuss prospects for planet habitability, as well as comment on the dynamical properties of the system. We also compare these planets to other known multi-planet systems to provide additional context for the TOI-700 system.

\subsection{Expected properties of TOI-700 e}

We used the \citet{forecaster} mass-radius relationship and the \texttt{forecaster} Python package to investigate the likely composition of TOI-700 e. With a radius of 0.953 R$_\oplus$, we found that TOI-700 e is likely a rocky planet with a probability of 87\%. \texttt{forecaster} yields a mass estimate of $0.845^{+0.67}_{-0.34}$ M$_\oplus$. Using this calculated mass, we estimate the expected RV semi-amplitude to be 37\,\cms. This small predicted signal will make a mass determination difficult, but robust upper mass limits are within reach of current instruments. A mass determination for TOI-700 e is further complicated by the fact that the orbital period (27.81 Days) is very close to  half that of the stellar rotation period \citep[$54\pm0.8$ days,][]{gilbert2020}. Following the methods of \citet{barnes08} and using the estimated mass of 0.845 M$_\oplus$, we calculated the timescale for tidal locking of TOI-700~e to be on order a few million years. Given the age of the system, it is likely that the planet is in a locked-in synchronous or pseudo-synchronous rotation.

\subsection{On the habitability of TOI-700 e}\label{sec:habitability}

\begin{figure*}
    \centering
    \includegraphics[width=\textwidth]{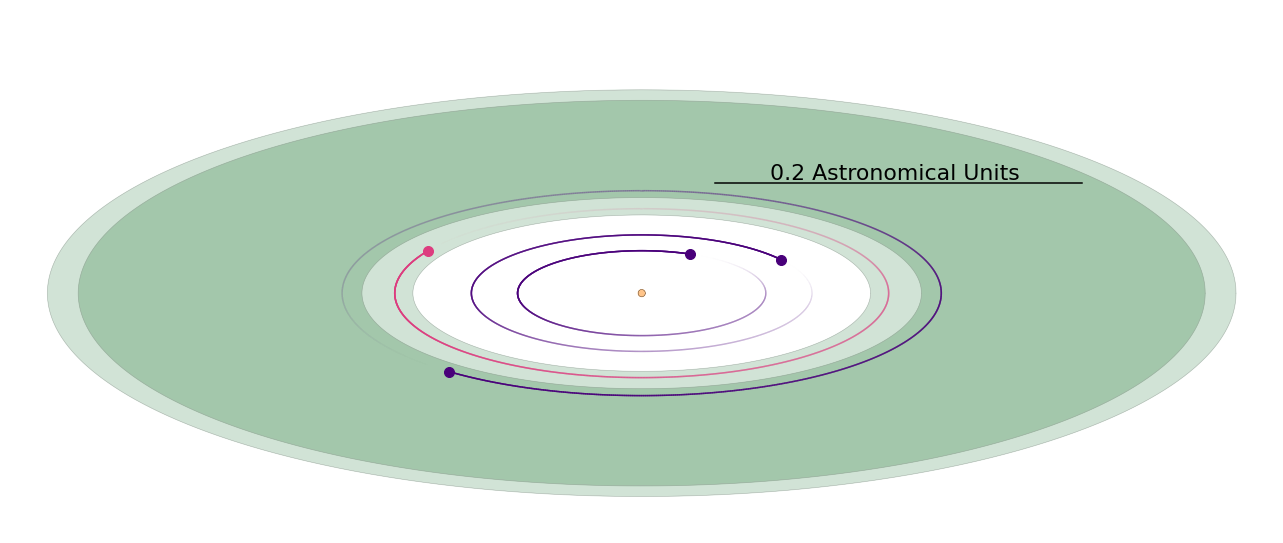}
    \caption{TOI-700 e (pink) resides in the Optimistic Habitable Zone (light green) around its host star in between the orbits of TOI-700 c and d. The Conservative Habitable Zone is shown in dark green, and planets b, c, and d (from inner to outer) are shown in indigo.}
    \label{fig:sys-overview}
\end{figure*}

The TOI-700 system is now known to contain 2 planets that orbit within the HZ of the host star, depicted in Figure~\ref{fig:sys-overview}. Specifically, we adopt the HZ boundaries provided by \citet{kasting1993a,kopparapu2013,kopparapu2014}, calculated using the stellar parameters from Table~\ref{tab:stellar-params}. The HZ is broadly divided into the Conservative HZ (CHZ) and the Optimistic HZ (OHZ), described in detail by \citet{kane2016c}. The CHZ boundaries are defined by the runaway and maximum greenhouse limits, and the OHZ boundaries are based upon empirically derived estimates of when Venus and Mars may have had surface liquid water. As shown in Figure~\ref{fig:sys-overview}, planet d resides within the CHZ 
($S_d = 0.85^{+0.09}_{-0.10}$\,\So; 0.246\,\So $<$ $S_{CHZ}$ $<$ 0.931\,\So) and the orbit of planet e is within the OHZ ($S_e = 1.27^{+0.13}_{-0.15}$\,\So; 0.221\,\So $<$ $S_{OHZ}$ $<$ 1.493\,\So). Systems such as TOI-700, with a potential diversity of terrestrial planetary climates, provide important context regarding climate evolution and comparative planetology \citep{kane2021a}. The insolation flux of TOI-700 e is $\sim$1.27\So, and thus lies between Earth and Venus ($\sim$1.91\So). Therefore, the presence of a terrestrial planet in the inner OHZ (planet e) and the CHZ (planet d) serve as potential analogs to the Venus/Earth evolutions that occurred within the solar system. The nature of the Venusian evolution remains a topic of some debate, with scenarios ranging from Venus never condensing surface liquid water \citep{hamano2013,turbet2021} to Venus retaining temperate conditions until as recently as a billion years ago \citep{way2016,way2020}. The TOI-700 system thus presents an opportunity to examine Venus within the exoplanet context \citep{kane2014,barstow2016a,kane2019d,lustigyaeger2019b,ostberg2019}.

Within the context of planetary habitability, it is also worth noting that the planetary parameters in Table~\ref{tab:planet-params} show that the estimated orbital eccentricities are relatively low, as expected for compact multi-transiting systems \citep{exoplanet:vaneylen19}. Eccentricities play an important role in the evolution and stability of terrestrial planet climates \citep{williams2002,dressing2010,kane2012e,kane2017d,way2017a}. The near-circular nature of the orbits increases the likelihood of stable climates and the long-term sustainability of temperate surface conditions.

\subsection{Stellar activity and candidate flares}

Another challenge to exoplanet habitability is the radiation environment. Low mass stars like TOI-700 are typically highly active for long portions of their lifetimes, producing frequent flares that can alter exoplanet atmospheres. TOI-700 was first observed at 2-minute cadence for 11 sectors during TESS Year 1 of observations. In all of these observations, there was no evidence of white-light flares in the light curves \citep{gilbert2020}. \citet{howard2022} searched for flares in the 2-minute cadence data of this target and also detected no flares.

In TESS Year 3 of observations, TOI-700 was observed for 10 full sectors of observations at 20-second cadence. This increased cadence allowed us to investigate short-lived phenomena that may not have been resolved in 2-minute cadence data. Upon investigating the 20-second cadence data, we identified 4 short-duration flare candidates that were not distinguishable in the 2-minute cadence data obtained at the same time. However, inspection of the raw 20 second data revealed that none of the candidates were flares, and most were instead caused by cosmic rays hitting the detector in multiple consecutive exposures\footnote{We note that for 20-second cadence data, cosmic ray mitigation is not performed on-board the spacecraft, and instead is performed in post-processing, so cosmic rays are often evident in the \texttt{raw\_cnts} images from the TESS target pixel files.}.

Additional observations will help us to further characterize the magnetic activity of TOI-700. Accurate characterization of the host star is necessary for our assessment of the habitability of the planets in this system. TESS will re-observe TOI-700 beginning in January 2023, providing us with 9 more sectors of observations that may be used to study any potential activity. Additional multi-wavelength and/or spectroscopic observations would also be valuable for characterizing the magnetic activity of this star. There was no evidence of flares in 15 orbits of HST observations (PI: Arney; Program: 16207) or 91 spectra taken with ESPRESSO (PI: Gilbert; Program: 108.22C2.001). These observations will be discussed in greater detail in future papers.

\subsection{Dynamical Analysis of the TOI-700 System}

TOI-700 e's orbit places it conspicuously close to the interior 4:3 mean motion resonance (MMR) with TOI-700 d ($P_{d}/P_{e}=$ 1.3458).  While modern theories of planet formation via pebble accretion and inward migration strongly suggest that compact systems of short-period planets form in chains of mutually resonant orbits \citep[e.g.][]{izidoro17}, proximity to a resonance does not imply resonant behavior exists in a system.  For instance, Jupiter and Saturn's orbital period ratio (2.49) is rather close to that of a 5:2 MMR, however the planets are not currently in this resonance, nor is it is likely that they ever were in the past \citep[e.g.][]{clement20}.  Confirmation of resonant evolution in exoplanet systems requires a more precise determination of the planets' angular orbital elements and their associated frequencies via precise radial velocity campaigns and transit timing variation \citep[TTVs, ][]{luger17,leleu21}.  In the absence of these data, we are forced to adopt a brute force-type of methodology that interrogates the stability of the 4:3 MMR within the TOI-700 system given the current measured uncertainties on the planets' orbits.
We performed a suite of 7,500 N-body simulations (with a duration of 10 Myr) of the 4 planet system using the \texttt{Mercury6} hybrid integrator \citep{chambers99}, a 0.5 day time-step, and additional forces to account for the effects of general relativity.  In each individual run, we assigned the nominal orbital parameters to planets b and c, randomly varied the semi-major axes of d and e within their cataloged uncertainties, and randomly assigned an eccentricity to the outer 2 planets between 0.0 and 0.05 by sampling from a uniform distribution of values.  While eccentricity is still not well constrained in this system, dynamical stability considerations \citep[][]{gilbert2020} strongly suggest that the eccentricities of all 4 planets are low.  In each iteration, we initialized the longitudes of perihelia and orbital anomalies of planets e and d in a configuration that placed them well within the libration island for the 4:3 MMR.  To determine whether each system evolved in MMR for the duration of the simulation, we checked for libration of the relevant resonant angles using the analysis pipeline described in \citet{clement21}.
In total, 521 of our 7,500 simulations (6.9$\%$) displayed the appropriate resonant libration.  However, when we restrict our sample to the 141 systems that were initialized with both planets e and d's eccentricities less than 0.01 (which is likely the case in this system), 54 simulations produced a resonance (38$\%$). We will be able to further assess this estimate with better constraints on the orbital parameters of the planets from ongoing radial velocity measurements and transit timing observations. While these simulations do not confirm the presence of a MMR in TOI-700, they are indeed a potential indication of the existence of one given the aforementioned formation models and the growing sample of similar systems discovered in resonant chains.  

The period ratio of 1.34 is not one that is produced in planet formation models without convergent migration and resonance capture \citep[gas-free planetesimal accretion models tend to produce period ratios of $\sim$1.6-2.5,][]{chambers98}, so it is likely that planets e and d formed in resonance as the result of orbital migration. While the system may have formed in resonance and subsequently destabilized, given the compact nature of this system and the current period ratios of the planets, it is unlikely to have experienced a scattering event under instability \citep{izidoro17, izidoro21}. Given these results, we conclude that it is quite probable that a MMR exists between planets e and d.  We plan to leverage on-going and planned observations of the system in a forthcoming manuscript to test this assertion.

We originally explored the possibility of TTVs in the system in \citet{gilbert2020} using a photodynamical model implemented using \texttt{TRANSITFIT5} \citep{Rowe2015,Rowe2016c} using the first year of TESS observations. Building on this analysis, we explored the possibility of TTVs within this system by analyzing the additional year of observations with the new planet, TOI-700 e, using the \texttt{TTVFast} Python package. We modeled a variety of conditions for all 4 planets, sampling over a range of orbital eccentricities and planet masses. In 1,000 realizations of the system, planets b and c only show small TTVs at most (on order a few minutes). Planets e and d show potentially larger variations (TTVs of up to a few hours), as a result of near resonant TTV behavior if eccentricity is as high as 0.05, but we do not see evidence of such large TTVs in the TESS photometry; we measured the TTVs for both planets, and they are consistent with 0 within errors, and have an RMS of 24 minutes for planet d, and 1 hour for planet e. This means that the eccentricities of the TOI-700 planets are likely very low, as one would expect for a compact, multi-planet system. Future transit observation with TESS and ongoing radial velocity measurements with ESPRESSO will help us to further constrain the orbital parameters of the planets in this system. These radial velocity measurements will also allow us to determine the masses of the TOI-700 planets, and place further constraints on expected TTV amplitudes.

\subsection{Comparison to other multi-planet systems}\label{sec:multi-planet}

The TESS mission has so far discovered over 100 planets in multi-planet systems, the majority of which orbit M dwarf stars. The L 98-59 (TOI-175) system was one of the earliest multi-planet systems discovered using TESS \citep{kostov2019b} and contains 3 terrestrial-sized, transiting planets \citep[and at least one non-transiting planet][]{Demangeon21}. L 98-59 is a bright ($V$ = 11.7 mag), nearby (10.6 pc) M3 dwarf that exhibits moderate amounts of activity in TESS optical light curves. L 231-32 (TOI-270) also hosts multiple transiting planets which were discovered in TESS data \citep{gunther19}. L 231-32 is also an M3 dwarf and hosts 3 super-Earth to sub-Neptune sized planets orbiting close to the star. L 98-59 was observed with HARPS \citep{cloutier19}, and both the L 98-59 and L 231-32 systems were subsequently followed up with ESPRESSO in order to obtain precise planet mass measurements \citep[][respectively]{Demangeon21,vaneylen21}. TESS also revealed the presence of 3 planets around M0 dwarf, TOI-1749 \citep{fukui2021}, ranging from super-Earth to sub-Neptune in size. These systems with multiple transiting planets are all likely coplanar given their geometries. Indeed, the inclinations of the TOI-700 planets are consistent with each other and are comparable to the inclinations of Solar System planets.

TOI-700 adds to the growing number of multi-planet systems discovered using TESS. Now with a second small, HZ planet known to orbit this star, this system is worthy of further investigation. Multi-planetary systems that harbor small planets in or near the HZ, \textit{and} orbit stars bright enough for follow-up atmospheric characterization, are quite rare. There are only a few dozen multi-planet systems that host temperate worlds. Even fewer of these worlds are small enough to be rocky or orbit stars bright enough for detailed follow-up observations.

Famously, TRAPPIST-1 hosts 7 known transiting planets \citep{Gillon2017}, several of which reside within the star's HZ. Comparing between the TOI-700 and TRAPPIST-1 systems with their multiple small, HZ planets provides us with valuable insight into what processes affect planetary habitability besides simply insolation flux. We are able to compare between host stars on either side of the convective boundary, as well as host stars that are quiet (TOI-700) vs. active (TRAPPIST-1). In these systems with multiple HZ planets, we know that the planets' host stars are the same, so their evolutionary histories are well controlled, and any differences in habitability can be attributed to phenomena other than the star's radiation environment or evolution. 

\subsection{Prospects for Future Investigations}
We conducted a transit search on our reduced light curves (described in Section \ref{sec:tess-observations}) and did not identify any additional signals (beyond the ones already detected) that may be indicative of other planets in the system. However, TESS will re-observe TOI-700 for 9 sectors in its upcoming Cycle 5 observations. Additional photometry may reveal additional low signal planets in this system, as could ongoing ESPRESSO precision radial velocity measurements (Programs 108.22C2.001 and 110.242P.001, PI: Gilbert), which will be discussed in a forthcoming paper (Gilbert et al., in prep).

In the absence of extracted RV masses at this point in time, we used \citet{forecaster} mass estimates to calculate the transmission spectroscopy metric \citep[TSM;][]{Kempton2018} for each of the TOI-700 planets to be 4 (planet b), 76 (planet c), 4 (planet e), and 3 (planet d). For reference, TOI-700 e receives an insolation flux 1.25 times that of Earth's irradiance, which is between TRAPPIST-1 c (2.27 S$_\oplus$) and TRAPPIST-1 d (1.143 S$_\oplus$), \citep{Gillon2017}. TRAPPIST-1 c and d have significantly higher TSM values (26 and 24 respectively) because of the comparatively smaller size of their host star. However, it is worth revisiting these metrics once we have precisely measured masses for the TOI-700 planets to fully assess their observability.

\section{Conclusions}\label{sec:conclusion}

We present the discovery of a new, small planet orbiting within the HZ of TOI-700 called TOI-700 e. TOI-700 e is a $0.95$ R$_\oplus$ planet which orbits its host star every 27.8 days. Due to its small size, this planet was not detected in Year 1 of TESS observations alone, but data from the TESS extended mission revealed its existence. TOI-700 e orbits interior to TOI-700 d \citep{gilbert2020, Rodriguez2020} and receives 1.27$\times$ Earth's insolation flux (as compared to 0.86 for TOI-700 d), putting it within the star's Optimistic HZ. TOI-700 e has a low-eccentricity orbit, likely rotates in a synchronized state with its 27.8 day orbit, and may be in a mean motion resonance with its fellow Earth-sized, HZ planet, TOI-700 d. 

TOI-700 is a notable system because it hosts multiple small planets, 2 of which are now known to orbit within the star's Habitable Zone. Most importantly, TOI-700 is relatively quiet and bright enough that we can now take the next steps to further characterize this system and learn more about its HZ planets. This lack of activity is also greatly beneficial in terms of planetary habitability. We see no signs of flares in the TESS optical photometry and TOI-700 is quiet when observed by HST in the UV, which is valuable in terms of planets' abilities to retain their atmospheres over time. TOI-700 will present an important contrast to other systems such as TRAPPIST-1 that host small, HZ planets, but have higher levels of stellar activity. A comparison between these planets may elucidate how the stellar environment affects planets' ability to maintain their atmospheres and what their compositions are like. The more systems we can study with multiple HZ planets, the better our understanding of these worlds will be.

\section*{Acknowledgements}

This paper makes use of the 20-second cadence mode introduced in the TESS extended mission. We thank those on the TESS team who made this data collection mode possible.

This paper includes data collected by the TESS mission, which are publicly available from the Mikulski Archive for Space Telescopes (MAST) at the Space Telescope Science Institute. The specific observations analyzed can be accessed via \dataset[doi:10.17909/t9-tcn7-7g94]{https://doi.org/doi:10.17909/t9-tcn7-7g94}, \dataset[doi:10.17909/t9-yk4w-zc73]{https://doi.org/doi:10.17909/t9-yk4w-zc73}, and \dataset[doi:10.17909/t9-yjj5-4t42]{https://doi.org/doi:10.17909/t9-yjj5-4t42}.

Funding for the TESS mission is provided by NASA's Science Mission directorate. We acknowledge the use of public TESS Alert data from pipelines at the TESS Science Office and at the TESS Science Processing Operations Center.

This research has made use of the Exoplanet Follow-up Observation Program website, which is operated by the California Institute of Technology, under contract with the National Aeronautics and Space Administration under the Exoplanet Exploration Program. 

This research was carried out in part at the Jet Propulsion Laboratory, California Institute of Technology, under a contract with the National Aeronautics and Space Administration (80NM0018D0004). 

The dynamical computations presented here were supported by the Carnegie Institution and were conducted in the Resnick High Performance Computing Center, a facility supported by Resnick Sustainability Institute at the California Institute of Technology.

Resources supporting this work were provided by the NASA High-End Computing (HEC) Program through the NASA Advanced Supercomputing (NAS) Division at Ames Research Center for the production of the SPOC data products.

E.A.G. thanks the LSSTC Data Science Fellowship Program, which is funded by LSSTC, NSF Cybertraining Grant \#1829740, the Brinson Foundation, and the Moore Foundation; her participation in the program has benefited this work. E.A.G. is thankful for support from GSFC Sellers Exoplanet Environments Collaboration (SEEC), which is funded by the NASA Planetary Science Division’s Internal Scientist Funding Model. The material is based upon work supported by NASA under award number 80GSFC21M0002. This work was also supported by NASA awards 80NSSC19K0104 and 80NSSC19K0315.

M.N.G. acknowledges support from the European Space Agency (ESA) as an ESA Research Fellow.

B.J.H. acknowledges support from the Future Investigators in NASA Earth and Space Science and Technology (FINESST) program grant 80NSSC20K1551.

C.H. thanks the support of the ARC DECRA program DE210101893.

The team thanks Laura Kreidberg and Ravi Kopparapu for their contributions to the TESS GI proposal. Authors 2 and 3 contributed equally, and the order was decided by the best performance in picking games against the spread of the NFL’s 2022 season week 3.

© 2022. All rights reserved.

\facilities{CAO, TESS}

\software{
DAVE \citep{Kostov2019}, 
exofast \citep{Eastman13,Eastman:2019}
Forecaster \citep{forecaster}, 
Jupyter \citep{jupyer}, 
Lightkurve \citep{lightkurve}, 
Matplotlib \citep{matplotlib},
Mercury6 \citep{chambers99},
NumPy \citep{numpy}, 
Pandas \citep{pandas}, 
TTVFast \citep{Deck2014}, 
vespa \citep{Morton2012,vespa},
xoflares \citep{xoflares}
}

\bibliography{refs}{}
\bibliographystyle{aasjournal}

\end{document}

%% file: updated-exofast-table.tex
\begin{deluxetable*}{l l c c c}[ht]
\tablecaption{Median values and 68\% confidence intervals for the global models for the TOI-700 stellar parameters.\label{tab:stellar-params}}
\tablewidth{0pt}
\tablehead{
\colhead{Parameter} & \colhead{Description (Units)}  & \colhead{Median} & \colhead{+1$\sigma$}  & \colhead{-1$\sigma$}
}
\startdata
M$_*$\dotfill & Mass (M$_\odot$)\dotfill &  0.415 & 0.021 & 0.020\\
R$_*$\dotfill & Radius (R$_\odot$)\dotfill & 0.421 & 0.017 & 0.015\\
L$_*$\dotfill & Luminosity (L$_\odot$)\dotfill & 0.0229 & 0.0026 & 0.0023\\
$\rho_*$\dotfill & Density (cgs)\dotfill & 7.86 & 0.91 & 0.88\\
$\log{g}$\dotfill & Surface gravity (cgs)\dotfill & 4.809 & 0.034 & 0.037\\
T$_{\rm eff}$\dotfill &Effective Temperature (K)\dotfill & 3459 & 65 & 65\\
$[{\rm Fe/H}]$\dotfill & Metallicity (dex)\dotfill &-0.07 & 0.11 & 0.11 \\
\enddata
\end{deluxetable*}

%% file: planet-params.tex
\begin{deluxetable*}{l l c c c}[ht]
\tablecaption{Median values and 68\% confidence intervals for the global models of the TOI-700 planetary parameters.\label{tab:planet-params}}
\tablewidth{0pt}
\tablehead{
\colhead{Parameter} & \colhead{Description (Units)}  & \colhead{Median} & \colhead{+1$\sigma$}  & \colhead{-1$\sigma$}
}
\startdata
\textbf{TOI-700 b}\\
P\dotfill &Period (days)\dotfill & 9.977219 & 0.000041 & 0.000038\\
R$_{P}$\dotfill &Radius (R$_\oplus$)\dotfill & 0.914 & 0.053 & 0.049\\
T$_0$\dotfill & Optimal conjunction Time (\bjdtdb)\dotfill & 2458880.0993 & 0.0015 & 0.0016\\
a\dotfill &Semi-major axis (AU)\dotfill & 0.0677 & 0.0011 & 0.0011\\
i\dotfill &Inclination (Degrees)\dotfill & 89.60 & 0.27 &0.29\\
e\dotfill &Eccentricity \dotfill & 0.075 & 0.093 & 0.054\\
$\omega_*$\dotfill &Argument of Periastron (Degrees)\dotfill & -150 & 120 &130\\
S\dotfill &Insolation Flux (S$_\oplus$)\dotfill & 4.98 & 0.50 & 0.58\\
$R_P/R_*$\dotfill &Radius of planet in stellar radii \dotfill & 0.01993 & 0.00082 & 0.00083\\
$T_{14}$\dotfill &Total transit duration (days)\dotfill & 0.0905 & 0.0036 & 0.0033\\
b\dotfill &Transit Impact parameter \dotfill & 0.24 & 0.19 & 0.16\\
\textbf{TOI-700 c}\\
P\dotfill &Period (days)\dotfill & 16.051137 & 0.000020 & 000020\\
R$_{P}$\dotfill &Radius (R$_\oplus$)\dotfill & 2.60 & 0.14 & 0.13\\
T$_0$\dotfill & Optimal conjunction Time (\bjdtdb)\dotfill & 2458821.62181 & 0.00041 & 0.00042\\
a\dotfill &Semi-major axis (AU)\dotfill & 0.0929 & 0.0015 & 0.0015\\
i\dotfill &Inclination (Degrees)\dotfill & 88.903 & 0.071 & 0.087\\
e\dotfill &Eccentricity \dotfill & 0.068 & 0.070 & 0.049\\
$\omega_*$\dotfill &Argument of Periastron (Degrees)\dotfill & 30 & 120 & 120\\
S\dotfill &Insolation Flux (S$_\oplus$)\dotfill & 2.64 & 0.26 & 0.31\\
$R_P/R_*$\dotfill &Radius of planet in stellar radii \dotfill & 0.0565 & 0.0017 & 0.0016\\
$T_{14}$\dotfill &Total transit duration (days)\dotfill & 0.0593 & 0.0017 & 0.0017\\
b\dotfill &Transit Impact parameter \dotfill & 0.899 & 0.014 & 0.017\\
\textbf{TOI-700 e}\\
P\dotfill &Period (days)\dotfill & 27.80978 & 0.00046 & 0.00040\\
R$_{P}$\dotfill &Radius (R$_\oplus$)\dotfill & 0.953 & 0.089 & 0.075\\
T$_0$\dotfill & Optimal conjunction Time (\bjdtdb)\dotfill & 2458964.8112 &0.0058 &0.0045\\
a\dotfill &Semi-major axis (AU)\dotfill & 0.1340 & 0.0022 & 0.0022\\
i\dotfill &Inclination (Degrees)\dotfill & 89.60 & 0.21 & 0.16\\
e\dotfill &Eccentricity \dotfill & 0.059 & 0.057 & 0.042\\
$\omega_*$\dotfill &Argument of Periastron (Degrees)\dotfill & 70 & 100 & 120\\
S\dotfill &Insolation Flux (S$_\oplus$)\dotfill & 1.27 & 0.13 & 0.15\\
$R_P/R_*$\dotfill & Radius of planet in stellar radii \dotfill & 0.0208 & 0.0017 & 0.0014\\
$T_{14}$\dotfill &Total transit duration (days)\dotfill & 0.1157 & 0.0087 & 0.016\\
b\dotfill &Transit Impact parameter \dotfill & 0.47 & 0.19 & 0.25\\
\textbf{TOI-700 d}\\
P\dotfill &Period (days)\dotfill & 37.42396 & 0.00039 & 0.00035\\
R$_{P}$\dotfill &Radius (R$_\oplus$)\dotfill & 1.073 & 0.059 & 0.054\\
T$_0$\dotfill & Optimal conjunction Time (\bjdtdb)\dotfill & 2458816.9952 & 0.0011 & 0.0013\\
a\dotfill &Semi-major axis (AU)\dotfill & 0.1633 & 0.0027 & 0.0027\\
i\dotfill &Inclination (Degrees)\dotfill & 89.80 & 0.12 & 0.10\\
e\dotfill &Eccentricity \dotfill & 0.042 & 0.045 & 0.030\\
$\omega_*$\dotfill &Argument of Periastron (Degrees)\dotfill &10 & 120 & 140\\
S\dotfill &Insolation Flux (S$_\oplus$)\dotfill & 0.85 & 0.09 & 0.10\\
$R_P/R_*$\dotfill & Radius of planet in stellar radii \dotfill & 0.02339 & 0.00077 & 0.00080\\
$T_{14}$\dotfill &Total transit duration (days)\dotfill & 0.1381 & 0.0028 & 0.0023\\
b\dotfill &Transit Impact parameter \dotfill & 0.29 & 0.14 & 0.18\\
\enddata
\end{deluxetable*}